\documentclass[aps,pra,twocolumn,superscriptaddress,nofootinbib]{revtex4-1}
\usepackage{amsmath}
\usepackage{amssymb}
\usepackage{graphicx}
\usepackage{mathrsfs}
\usepackage{ntheorem}
\usepackage{mathtools}
\usepackage{enumerate}
\usepackage{enumitem}
\usepackage{times,txfonts}
\usepackage{subfigure}
\usepackage{upgreek}
\usepackage{tikz}
\usepackage[colorlinks,bookmarksopen,bookmarksnumbered,citecolor=blue, linkcolor=blue, urlcolor=blue]{hyperref}
\newcommand{\ket}[1]{|#1\rangle}
\newcommand{\bra}[1]{\langle #1|}
\newcommand{\Tr}{\mathrm{Tr}}

\newcommand{\abs}[1]{\lvert #1\rvert}

\def\CC{{\rm\kern.24em \vrule width.04em height1.46ex depth-.07ex \kern-.30em C}}
\def\RR{{\rm\kern.24em \vrule width.04em height1.46ex depth-.07ex
\kern-.30em R}}
\def\P{{\rm I\kern-.25em P}}

\begin{document}
\title{Closed expressions for one-qubit states of convex roof coherence measures \\}

\author{Xiao-Dan Cui}
\affiliation{Shandong Provincial Engineering and Technical Centre of Light Manipulations and Shandong Provincial
Key Laboratory of Optics and Photonic Device, School of Physics and Electronics, Shandong Normal University, Jinan 250100, China}

\author{C. L. Liu}
\email{clliusdu@foxmail.com}
\affiliation{Graduate School of China Academy of Engineering Physics, Beijing 100193, China}

\date{\today}

\begin{abstract}
We study the closed expressions of the convex roof coherence measures for one-qubit states in this paper. We present the analytical expressions for the convex roof coherence measures, $C_f(\rho)$, of one-qubit states with $C_f(\varphi):=f(\abs{c_0}^2,\abs{c_1}^2)$ (where $\ket{\varphi}=c_0\ket{0}+c_1\ket{1}$) being convex with respect to the $l_1$ norm of coherence of $\varphi$ (i.e., $C_{l_1}(\varphi)$), such coherence measures including the coherence of formation, the geometric measure of coherence, the coherence concurrence, and the coherence rank. We further present the operational interpretations of these measures. Finally, we present the usefulness of the convex roof coherence measures $C_f(\varphi)$ being non-convex with respect to $C_{l_1}(\varphi)$ by giving the necessary and sufficient conditions for the transformations $p\varphi_1\oplus(1-p)\varphi_2\to q\phi_1\oplus(1-q)\phi_2$ via incoherent operations, where $\varphi_i$, $\phi_j$ $(i, j=1, 2)$ are one-qubit pure states and $0\leq p, q\leq 1$.
\end{abstract}
\maketitle

\section{Introduction}
Quantum coherence, which is a fundamental feature of quantum mechanics, represents a useful resource for performing various quantum information processing tasks with broad applications in a plethora of fields, ranging from quantum algorithms and quantum cryptography \cite{Nielsen}, to nanoscale thermodynamics \cite{Lostaglio1}, quantum metrology \cite{Giovannetti}, and quantum biology \cite{Lloyd,Huelga}. With the development of quantum information science, much attention has been paid to the quantification of coherence \cite{Aberg,Levi,Baumgratz,Streltsov,Fan,Wu}.

By adopting the viewpoint of coherence as a physical resource, Baumgratz \textit{et al.} proposed a seminal framework for quantifying coherence \cite{Baumgratz}. In that framework, a functional of states can be taken as a coherence measure if it fulfills four conditions, namely, the coherence being zero (positive) for incoherent states (all other states), the monotonicity of coherence under incoherent operations, the monotonicity of coherence under selective measurements on average, and the nonincreasing of coherence under mixing of quantum states.

By following the framework, a great number of coherence measures have been proposed \cite{Streltsov,Fan,Wu,Cui,Liu4}, some of which are defined based on distance, such as the $l_1$ norm of coherence and the relative entropy of coherence \cite{Baumgratz}, while others are defined based on the convex roof construction, such as the geometric measure of coherence \cite{Streltsov1}, the coherence of formation \cite{Aberg,Yuan}, and the coherence concurrence \cite{Qi}. These measures have been widely used to address various topics on quantum coherence \cite{Bromley,Yu1,Bu,Meng,Yuan,Winter,Liu2,Regula1,Liu,LiuZ,Lemi,Qi,Streltsov1,Yao,Singh,Radhakrishnan,Ma,Chitambar2,Tan,Guo,Liu3,Zhang}, such as the dynamics of quantum coherence \cite{Bromley,Yu1,Bu,Meng}, the distillation of quantum coherence \cite{Yuan,Winter,Liu2,Regula1,Liu,LiuZ,Lemi}, and the relations between quantum coherence and other quantum resources \cite{Qi,Streltsov1,Yao,Singh,Radhakrishnan,Ma,Chitambar2,Tan,Guo,Liu3}.

Since convex roof coherence measures involve an optimization process, it is generally not possible to directly present closed expressions for them \cite{Streltsov}. Although closed expressions for some specific convex roof coherence measures for one-qubit states have been evaluated, closed expressions for general convex roof coherence measures have never been rigorously provided before.

In this paper, we will study the problem mentioned above. Specifically, we will investigate the closed expressions of the convex roof coherence measures for one-qubit states. Our findings can be summarized as follows:
(1) we have obtained analytical expressions for the convex roof coherence measures, $C_f(\rho)$, for one-qubit states, where $C_f(\varphi)$ is continuous and convex with respect to $C_{l_1}(\varphi)$;
(2) we have obtained analytical expressions for the convex roof coherence measures, $C_f(\rho)$, for one-qubit states, where $C_f(\varphi)$ is discontinuous and convex with respect to $C_{l_1}(\varphi)$, such as the coherence rank \cite{Winter};
(3) although we cannot derive analytical expressions for the convex roof coherence measures that are non-convex with respect to $C_{l_1}(\varphi)$, we present their operational interpretation by providing the necessary and sufficient conditions for the transformation
\begin{eqnarray}
p\varphi_1\oplus(1-p)\varphi_2\to q\phi_1\oplus(1-q)\phi_2 \label{transformation1}
\end{eqnarray}
to be achieved using incoherent operations. Here, $\varphi_i$, $\phi_j$ with $i, j=1, 2$ are one-qubit pure states, and $0\leq p, q\leq 1$.

This paper is organized as follows: In Sec. \ref{II}, we will review some of the key concepts of the quantum resource theory of coherence. In Sec. \ref{III}, we will present the analytical expressions of the convex roof coherence measures $C_f(\rho)$ for one-qubit states, where $C_f(\varphi)$ is convex with respect to $C_{l_1}(\varphi)$. In Sec. \ref{IV}, we will discuss the operational interpretations of the convex roof coherence measures. Finally, in Sec. \ref{V}, we will present our conclusions.

\section{Resource theory of coherence} \label{II}

Let $\mathcal{H}$ represent the Hilbert space of a $d$-dimensional quantum system. A particular basis of $\mathcal{H}$ is denoted as $\{\ket{i},i=0,1,\ldots,d-1\}$, which is chosen according to the physical problem under consideration. The coherence of a state is then measured based on the chosen basis. We use $\rho=\sum_{ij}\rho_{ij}\ket{i}\bra{j}$ to denote a general density operator in the basis, where $\rho_{ij}$ are the elements of the density matrix. A state is called incoherent if its density operator is diagonal in the basis, and the set of all incoherent states is denoted by $\mathcal{I}$. It follows that a density operator $\rho$ belonging to $\mathcal{I}$ takes the form $\rho=\sum^{d-1}_{i=0}\rho_{ii}\ket{i}\bra{i}$. All other states, which cannot be written as diagonal matrices in the basis, are called coherent states. A general pure state is denoted by $\ket{\varphi}=\sum_{i=0}^{d-1} c_i\ket{i}$ with $c_i$ being the coefficients, corresponding to the density operator $\varphi=\ket{\varphi}\bra{\varphi}$.

An incoherent operation is defined by a completely positive and trace-preserving (CPTP) map, $\Lambda(\rho)=\sum_n K_n\rho K_n^\dagger$, where the Kraus operators satisfy two conditions: $\sum_n K_n^\dagger K_n= I$ and $K_n\mathcal{I}K_n^\dagger\subset \mathcal{I}$. This means that each $K_n$ maps an incoherent state to another incoherent state.

A functional $C$ can be considered as a coherence measure if it satisfies the following four conditions \cite{Baumgratz}:
\begin{itemize}
\item[C1] $C(\rho)\ge 0$, and $C(\rho)=0$ if and only if $\rho\in\mathcal{I}$;
\item[C2] Monotonicity under incoherent operations, $C(\rho)\ge C(\Lambda(\rho))$ if $\Lambda$ is an incoherent operation;
\item[C3] Monotonicity under selective measurements on average, $C(\rho)\ge \sum_np_nC(\rho_n)$, where $p_n=\Tr(K_n\rho K_n^\dagger)$,
$\rho_n=K_n\rho K_n^\dagger/p_n$, and $\Lambda(\rho)=\sum_nK_n\rho K_n^\dagger$ is an incoherent operation;
\item[C4] Non-increasing under mixing of quantum states, i.e., convexity, $\sum_nq_nC(\rho_n)\ge C(\sum_nq_n\rho_n)$ for any set of states $\{\rho_n\}$ and any probability distribution $\{q_n\}$.
\end{itemize}

Based on the framework for quantifying coherence, a great number of coherence measures have been proposed \cite{Streltsov,Wu,Fan}. Here, we recall a coherence measure, the $l_1$ norm of coherence \cite{Baumgratz}, which will be used in this paper. If we use $\rho=\sum_{i,j=0}^{d-1}\rho_{ij}\ket{i}\bra{j}$ to represent a general state, the $l_1$ norm of coherence is defined straightforwardly by the sum of absolute values of all the off-diagonal elements,
\begin{eqnarray}
C_{l_1}(\rho)=\sum_{i\neq j}|\rho_{ij}|.\label{L}
\end{eqnarray}
Another main set of coherence measures is the convex roof coherence measures. For a pure state $\ket{\varphi}=\sum_{i=0}^{d-1}c_i\ket{i}$, we define $C_f(\varphi):=f(|c_0|^2,|c_1|^2,...,|c_{d-1}|^2)$. The extension to mixed states is accomplished using the standard convex roof construction \cite{Du, Zhu}. This measure can be expressed in general as follows:
\begin{equation}
C_f(\rho) = \inf_{\{p_i, \varphi_i\}} \sum_i p_i C_f(\varphi_i), \label{convex roof}
\end{equation}
where the infimum is taken over all possible ensemble decompositions $\rho = \sum_i p_i\ket{\varphi_i}\bra{\varphi_i}$ with $p_i \geq 0$ and $\sum_i p_i = 1$.

\section{Closed expressions for one-qubit convex roof coherence measures} \label{III}
Equipped with the above notions, we show the analytical expression of the convex roof coherence measures for one-qubit states with $C_f(\varphi)$ being convex with respect to $C_{l_1}(\varphi)$.

\emph{Theorem 1.} Let $\ket{\varphi}=c_0\ket{0}+c_1\ket{1}$ with $\abs{c_0}\leq\abs{c_1}$. Suppose there exists a function $\hat{f}\left(\abs{c_0c_1^*}\right):=f\left(\abs{c_0}^2,\abs{c_1}^2\right)$ that is continuous and convex with respect to $\abs{c_0c_1^*}$, such that
\begin{eqnarray}
C_f(\varphi)=\hat{f}\left(\abs{c_0c_1^*}\right).
\end{eqnarray}
Then, for any arbitrary one-qubit state $\rho=\sum_{i,j=0}^1\rho_{ij}\ket{i}\bra{j}$, we have
\begin{eqnarray}
C_f(\rho)=\hat{f}\left(\abs{\rho_{01}}\right).
\end{eqnarray}

\emph{Proof.} According to \cite{Du1}, the set of one-qubit pure states is a totally ordered set, which implies that the function $\hat{f}$ is monotonically increasing. Suppose we have an optimal ensemble decomposition $\{p_i,\varphi_i\}$ satisfying the equality in Eq. (\ref{convex roof}), i.e., $C_f(\rho)=\sum_ip_i C_f(\varphi_i)$. Then, we can show that
\begin{eqnarray}
C(\rho)=\sum_ip_iC(\varphi_i)=\sum_ip_i\hat{f}\left(\abs{c_i^0 c_i^{1*}}\right)
\nonumber\\
\geq \hat{f}\left(\sum_ip_i\abs{c_i^0 c_i^{1*}}\right)\geq \hat{f}\left(\abs{\rho_{01}}\right),\label{theorem1}
\end{eqnarray}
where $\ket{\varphi_i}:=c_i^0\ket{0}+c_i^1\ket{1}$, the first inequality follows from the convexity of the function $\hat{f}$, and the last inequality is a consequence of the triangle inequality and the fact that $\hat{f}$ is monotonically increasing.

Next, we will prove that for any one-qubit state $\rho$, there exists an ensemble of $\rho$ that satisfies all the equalities in Eq. (\ref{theorem1}). To do this, let us show that $C(\rho)=C(\rho^\prime)$ for all coherence measures, where
\begin{eqnarray}
\rho=
\left(
    \begin{array}{cc}
        \rho_{00} & \rho_{01}\\
        \rho_{10} & \rho_{11}\\
    \end{array}
\right)
\end{eqnarray}
and
\begin{eqnarray}
\rho^\prime=
\left(
    \begin{array}{cc}
        \rho_{00} & \abs{\rho_{01}}\\
        \abs{\rho_{10}} & \rho_{11}\\
    \end{array}
\right).
\end{eqnarray}
We notice that there are $\rho=U^{\dagger}\rho^\prime U$ and $U\rho U^{\dagger}=\rho^\prime$ with $U:=\textrm{diag}(1,e^{i\arg(\rho_{01})})$ being an incoherent unitary operator. Thus, $C(\rho)=C(\rho^\prime)$ follows immediately by using condition (C2). Then, we will show that there exists an ensemble $\{p^\prime_i,\varphi^\prime_i\}$ of $\rho^\prime$ such that $C(\varphi^\prime_i)=\hat{f}(\abs{\rho_{01}})$ for each $i$. To this end, we introduce two pure states defined by
$$\ket{\varphi^\prime_1}=\sqrt{q}\ket{0}+\sqrt{1-q}\ket{1}$$
and
$$\ket{\varphi^\prime_2}=\sqrt{1-q}\ket{0}+\sqrt{q}\ket{1}$$
respectively, where $q$ is a non-negative number satisfying $\sqrt{q(1-q)}=\abs{\rho_{01}}$. Since $\sqrt{q(1-q)}=\abs{\rho_{01}}\leq\sqrt{\rho_{00}\rho_{11}}=\sqrt{\rho_{00}(1-\rho_{00})}$, we have that $\rho_{00}$ lies between $q$ and $(1-q)$. Hence, there exists a number $0\leq p^\prime\leq 1$ such that $\rho_{00}=p^\prime q+(1-p^\prime)(1-q)$, where $\rho_{00}\leq\rho_{11}$. Direct calculations show that $\rho^\prime=p^\prime\ket{\varphi^\prime_1}\bra{\varphi^\prime_1}+(1-p^\prime)\ket{\varphi^\prime_2}\bra{\varphi^\prime_2}$ and $C(\varphi^\prime_1)=C(\varphi^\prime_2)=\hat{f}(\abs{\rho_{01}})$. Thus, we arrive at the desired ensemble of $\rho^\prime$. ~~~~~~~~~~~~~~~~~~~~~~~~~~~~~~~~~~~~~~~~~~~~~~~~~~~~~~~~~~~~~~~~~~~~~~~~~~~~~~~~~~~~ $\blacksquare$

For a pure state $\ket{\varphi}=c_0\ket{0}+c_1\ket{1}$, we have $C_{l_1}(\varphi)=2\abs{c_0c_1^*}$. Consequently, if $C(\rho)$ is a continuous convex roof coherence measure and $C_f(\varphi)=\check{f}(C_{l_1}(\varphi))$ is convex with respect to  $C_{l_1}(\varphi)$ for some $\check{f}$, then $C_f(\rho)=\check{f}(C_{l_1}(\rho))$ holds for any one-qubit state. This theorem applies to various well-known convex roof coherence measures, such as the geometric measure of coherence \cite{Streltsov1}, the coherence of formation \cite{Aberg,Yuan}, and the coherence concurrence \cite{Qi}.

However, it is important to note that the equality in Eq. (\ref{theorem1}) does not always hold in general if $C_f(\varphi)$ is not continuous. A typical example is the coherence rank \cite{Winter}, which is defined as the number of non-zero terms in the decomposition of a pure state $\ket{\varphi}=\sum_{i=0}^{R-1}c_i\ket{i}$ with $c_i\neq 0$ minus 1, i.e., $C_R(\varphi)=R-1$ \cite{Liu1}, or as the logarithm of the number of non-zero terms in this decomposition, i.e., $C_R(\varphi)=\log R$ \cite{Xi}. For a mixed state $\rho$, its coherence rank is defined as $C_R(\rho)=\inf_{\{p_i,\varphi_i\}}\sum_ip_iC_R(\varphi_i)$, where $\rho=\sum_ip_i\ket{\varphi_i}\bra{\varphi_i}$ is any decomposition of $\rho$ into pure states $\varphi_i$ with $p_i\geq0$ \cite{Liu1,Xi}.

When considering the one-qubit case, the coherence rank is given by
\begin{eqnarray}\label{rank}
C_R(\rho)=\inf_{\{p_i,\varphi_i\}}\sum_{i}p_iC_R(\varphi_i),
\end{eqnarray}
where $C_R(\varphi)$ is defined for pure states $\ket{\varphi}$ as
\begin{equation}
C_ R(\varphi)=\left\{
  \begin{array}{ll}
    0,  ~~~~~~~~~ \ket{\varphi}\in \mathcal{I}\\
    1,  ~~~~~~~~~ \ket{\varphi}\notin \mathcal{I}
  \end{array}\right.
\end{equation}
with $\mathcal {I}$ being the set of incoherent states. We note that $C_R(\rho)$ identical to the coherence measure in Ref. \cite{Ding} in the one-qubit case. For these coherence measures, the following theorem holds.

\emph{Theorem 2.} Let $\rho=\sum_{i,j=0}^1\rho_{ij}\ket{i}\bra{j}$ be an arbitrary one-qubit state. Suppose there is $\rho_{00}\leq\rho_{11}$. Then
\begin{eqnarray}
C_R(\rho)= \left\{
  \begin{array}{ll}
    2\abs{\rho_{01}},  ~~~~~~~~~ &\mathrm{if}~\rho_{00}\geq \abs{\rho_{01}},\\
    \rho_{00}+\frac{\abs{\rho_{01}}^2}{\rho_{00}},  ~~~~~~~~~ &\mathrm{if}~\rho_{00}<\abs{\rho_{01}}.
  \end{array}\right.\label{theorem2}
\end{eqnarray}

\emph{Proof.} Let us divide $\rho$ into $\rho=\rho_c+\rho_i$ with $\rho_i$ being an (un-normalized) incoherent state, i.e., there are
\begin{eqnarray}
\rho_c=
\left(
    \begin{array}{cc}
        \rho^c_{00} & \rho_{01}\\
        \rho_{10} & \rho^c_{11}\\
    \end{array}
\right)
\end{eqnarray}
and
\begin{eqnarray}
\rho_i=
\left(
    \begin{array}{cc}
        \rho_{00}^i & 0\\
        0 & \rho_{11}^i\\
    \end{array}
\right)
\end{eqnarray}
with $\rho_{00}=\rho_{00}^c+\rho_{00}^i$. Suppose $\{p_i,\varphi_i\}$ is an optimal ensemble decomposition of $\rho$ that achieves the minimum in Eq. (\ref{rank}). Then, all the ensemble decompositions of $\rho_c$ have no incoherent elements, i.e., $\rho_c=\sum_i{C_R(\varphi_i)}p_i\varphi_i$. Thus it is direct to see that
\begin{eqnarray}
C_R(\rho)=\min \Tr(\rho_c).\label{optimal}
\end{eqnarray}
We show that the decomposition achieving the minimum in Eq. (\ref{optimal}) is obtained only when the coherence part $\rho_c$ is a pure state. We prove this by contradiction. To see this, suppose there is
\begin{eqnarray}
\rho=\rho_c+\rho_i=\sum_ip_i\ket{\varphi_i}\bra{\varphi_i}+\rho_i.   \label{decomposition}
\end{eqnarray}
Here, we first consider that $i=1,2$ and the generalization to other cases is straightforward. Without loss of generality, let
\begin{eqnarray}
\ket{\varphi_1}=c^0_1\ket{0}+c^1_1\ket{1}
\end{eqnarray}
and
\begin{eqnarray}
\ket{\varphi_2}=c^0_2\ket{0}+c^1_2\ket{1},
\end{eqnarray}
respectively. Then, there is $\rho_{01}^\prime=p_1c^0_1{c^1_1}^*+p_2c^0_2{c^1_2}^*=:\abs{\rho_{01}^\prime}e^{i\delta}$. Since $\rho_c$ is a mixed state, we immediately obtain that $\abs{\rho_{01}^\prime}^2<\rho^c_{00}\rho^c_{11}$. Without loss of generality, suppose $\rho^c_{00}\leq\rho^c_{11}$. Then there is $\tilde{\rho}_{11}^c:=\abs{\rho_{01}^\prime}^2/\rho^c_{00}<\rho^c_{11}$. Hence, we can always find a (un-normalized) pure state
\begin{eqnarray}
\ket{\hat{\varphi}}=\sqrt{\rho^c_{00}}\ket{0}+\sqrt{\tilde{\rho}_{11}^c}e^{i\delta}\ket{1}
\end{eqnarray}
such that the state $\rho$ has a decomposition
\begin{eqnarray}
\rho=\hat{\varphi}+\hat{\rho}_i,
\end{eqnarray}
where $\hat{\rho}_i:=\rho-\hat{\varphi}$ is an (un-normalized) incoherent state.
However, $\mathrm{Tr}(\hat{\varphi})=\rho^c_{00}+\tilde{\rho}_{11}^c<\mathrm{Tr}(\rho_c)=\rho^c_{00}+\rho_{11}^c$,
i.e., this decomposition has a smaller trace.
Thus, the contradiction implies that min $\mathrm{Tr}(\rho_c)$ can only be obtained when $\rho_c$ is a pure state.

Next, let us give the closed expression of min $\mathrm{Tr}(\rho_c)$. Since $\rho_c$ is a pure state, then there are $\rho_{00}^c\rho_{11}^c=\abs{\rho_{01}}^2$ and $\rho_{00}^c+\rho_{11}^c\geq2\sqrt{\rho_{00}^c\rho_{11}^c}=2\abs{\rho_{01}}$. Since we have assumed that $\rho_{00}<\frac{1}{2}$, we need to consider the following two cases: (i) $\rho_{00}<\abs{\rho_{01}}$ and (ii) $\rho_{00}\geq \abs{\rho_{01}}$.

In case (i), min $\mathrm{Tr}(\rho_c)$ can be obtained when $\rho_{00}^c=\rho_{00}$. Further, by direct calculations, we obtain that
\begin{eqnarray}
\rho_{11}^c=\frac{\abs{\rho_{01}}^2}{\rho_{00}},
\end{eqnarray}
and then
\begin{eqnarray}
\rho_{00}^c+\rho_{11}^c=\rho_{00}+\frac{\abs{\rho_{01}}^2}{\rho_{00}}.
\end{eqnarray}
Thus, we obtain
\begin{eqnarray}
C_R(\rho)=\min \mathrm{Tr}(\rho_c)
=\rho_{00}+\frac{\abs{\rho_{01}}^2}{\rho_{00}}.
\end{eqnarray}

In case (ii), min $\mathrm{Tr}(\rho_c)$ can be obtained when $\rho_{00}^c=\rho_{11}^c=\abs{\rho_{01}}$. In this case, we can obtain that
\begin{eqnarray}
C_R(\rho)=\min \mathrm{Tr}(\rho_c)=2\abs{\rho_{01}}.
\end{eqnarray}
This completes the proof of Theorem 2.~~~~~~~~~~~~~~~~~~~~~~~~~~~~~~~~~ $\blacksquare$

With the above theorems, we present some applications of the above theorems which are related to the state transformations under incoherent operations. For the coherence measures considered in Theorem 1, we have the following corollary:

\emph{Corollary} 1. Suppose $\varphi$ is a pure state of an one-qubit and $\rho$ is a mixed state of an one-qubit. The transformation from $\varphi$ to $\rho$ can be achieved by using an incoherent operation $\Lambda$ if and only if $C(\varphi)\geq C(\rho)$, where $C(\rho)$ is any coherence measure in Theorem 1.

\emph{Proof.} If there is $\Lambda(\varphi)=\rho$, then we have $C (\varphi)\geq C(\rho)$. On the other hand, if there is $C(\varphi)\geq C(\rho)$, then $\abs{\varphi_{01}}\geq\abs{\rho_{01}}$. Furthermore, since $\varphi$ is a pure state then there is $\abs{\varphi_{01}}^2=\varphi_{00}\varphi_{11}$ and since $\rho$ is a general state then there is $\abs{\rho_{01}}^2\leq\rho_{00}\rho_{11}$. By using a result in Ref. \cite{Chitambar1} which says that the transformation
\begin{eqnarray}
\rho=\left(
    \begin{array}{cc}
        \rho_{00} & \rho_{01} \\
        \rho_{10}  & \rho_{11}\\
    \end{array}
\right)\to\sigma=\left(
    \begin{array}{cc}
        \sigma_{00} & \sigma_{01}\\
        \sigma_{10} & \sigma_{11}\\
    \end{array}
\right)
\end{eqnarray}
can be achieved using incoherent operations if and only if there are
\begin{eqnarray}
C_\zeta(\rho)\geq C_\zeta(\sigma)
\end{eqnarray}
and
\begin{eqnarray}
C_\xi(\rho)\geq C_\xi(\sigma), \label{condition}
\end{eqnarray}
where $C_\zeta(\rho)$ is defined as
\begin{eqnarray}
C_\zeta(\rho):=\abs{\rho_{01}}
\end{eqnarray}
and $C_\xi(\rho)$ is defined as
 \begin{eqnarray}
C_\xi(\rho):=\frac{\abs{\rho_{01}}}{\sqrt{\rho_{00}\rho_{11}}},
\end{eqnarray} we obtain that there is some incoherent operation $\Lambda$ such that $\Lambda(\varphi)=\rho$. ~~~~~~~~~~~~~~~~~~~~~~~~~~~~~~~~~~~~~~~~~~~~~~~~~~~~~~~~~~~~~~~~~~~~~~~~~~~~~~~~~~~$\blacksquare$

Next, we will present the operational interpretation of the coherence measures that are considered in Theorem 2. This leads to the following corollary:

\emph{Corollary} 2. Let $\rho$ be a mixed state of an one-qubit and let $\varphi$ be a pure coherent state of an one-qubit. It is impossible to transform $\rho$ into $\varphi$ using any incoherent operation.

\emph{Proof.} Let $\rho$ be a mixed state of an one-qubit. Then, we have $\abs{\rho_{01}}^2<\rho_{00}\rho_{11}$. Without loss of generality, we assume that $\rho_{00}\leq\rho_{11}$. Thus, we also have $\rho_{00}\leq\frac12$. If $\abs{\rho_{01}}\leq\rho_{00}<\frac12$ or $\abs{\rho_{01}}<\rho_{00}=\frac12$, then $C_R(\rho)=2\abs{\rho_{01}}<1$. If $\abs{\rho_{01}}=\rho_{00}=\frac12$, then $\rho$ is a pure state. Suppose $\abs{\rho_{01}}>\rho_{00}$. Then, we have $\rho_{00}<\abs{\rho_{01}}<\rho_{11}$. Using the inequality $\abs{\rho_{01}}^2<\rho_{00}\rho_{11}$, we obtain $C_R(\rho)=\rho_{00}+\frac{\abs{\rho_{01}}^2}{\rho_{00}}<\rho_{00}+\rho_{11}=1$. By contradiction, suppose there exists an incoherent operation $\Lambda$ such that $\Lambda(\rho)=\varphi$. Then, we have $C(\rho)\geq C(\varphi)$ for all coherence measures. However, this is not possible since $C_R(\rho)<C_R(\varphi)$. ~~~~$\blacksquare$

\section{Operational interpretation of $C_f(\varphi)$ being non-convex with respect to $C_{l_1}(\varphi)$}\label{IV}
In this section, we first demonstrate that the result in Theorem 1 cannot be directly applied to convex roof coherence measures where $C_f(\varphi)$ is non-convex with respect to $C_{l_1}(\varphi)$. We then provide an operational interpretation of these measures by presenting the necessary and sufficient conditions for Eq. (\ref{transformation1}) using incoherent operations.

\begin{figure}[h]
\vspace{-0mm}
\centering
\includegraphics[width=0.45\textwidth]{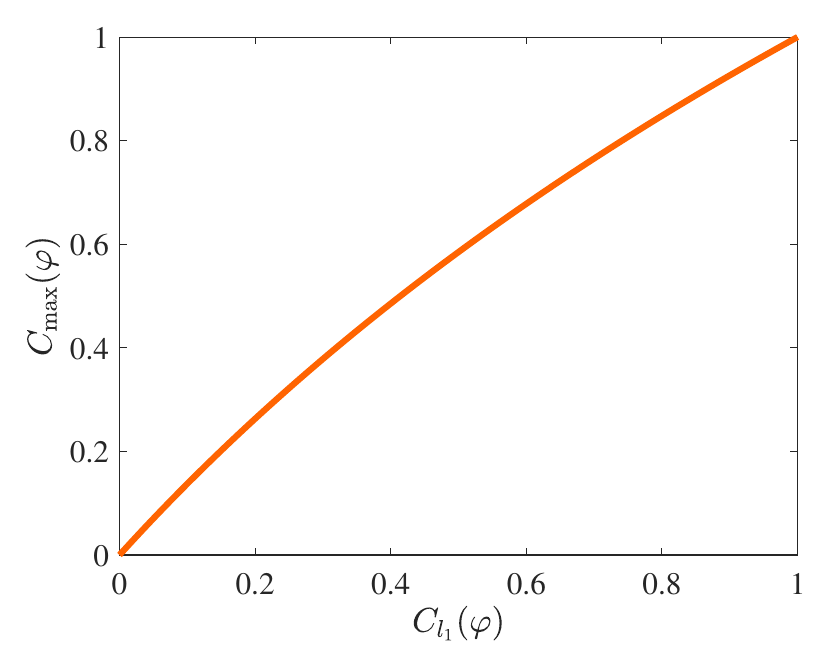}
\caption{The function $C_{\mathrm{max}}(\varphi)$ of $C_{l_1}(\varphi)$ with $\varphi$ being an one-qubit pure state. It is direct to see that $C_{\mathrm{max}}(\varphi)$ is a concave function with respect to $C_{l_1}(\varphi)$.}
\label{fig1}
\end{figure}

To this end, we first consider the convex roof of the maximum relative entropy of coherence \cite{Bu1}, which is defined as
\begin{eqnarray}
C_{\mathrm{max}}(\rho)=\inf_{\{p_i,\varphi_i\}}\sum_{i}p_iC_{\mathrm{max}}(\varphi_i).
\end{eqnarray}
Here, $C_{\mathrm{max}}(\rho):=\min_{\sigma\in\mathcal{I}}\min\{\uplambda|\rho\leq2^{\uplambda}\sigma\}$, and the infimum is taken over all ensembles $\{p_i,\varphi_i\}$ that realize $\rho$. It is straightforward to show that $C_{\mathrm{max}}(\varphi)=\log_2[1+C_{l_1}(\varphi)]$ \cite{Bu1}, and thus $C_{\mathrm{max}}(\varphi)$ is a concave function with respect to $C_{l_1}(\varphi)$ (see Fig. \ref{fig1}).
With the above results, we consider the state
\begin{eqnarray}
\rho=
\left(
    \begin{array}{cc}
        \frac{9}{32} & \frac{1}{4}+\frac{\sqrt{15}}{32}\\
        \frac{1}{4}+\frac{\sqrt{15}}{32} & \frac{23}{32}\\
    \end{array}
\right).
\end{eqnarray}
Let us consider an ensemble decomposition of $\rho=p\ket{\varphi_1}\bra{\varphi_1}+(1-p)\ket{\varphi_2}\bra{\varphi_2}$ with
\begin{eqnarray}
&p=\frac{1}{2},~~~\ket{\varphi_1}=\frac{1}{4}\ket{0}+\frac{\sqrt{15}}{4}\ket{1};
\nonumber\\
&1-p=\frac{1}{2},~~~\ket{\varphi_2}=\frac{1}{\sqrt{2}}\ket{0}+\frac{1}{\sqrt{2}}\ket{1}.
\end{eqnarray}
By direct calculations, we obtain $C_{\mathrm{max}}(\rho)\leq pC_{\mathrm{max}}(\varphi_1)+(1-p)C_{\mathrm{max}}(\varphi_2)=\log_2(\frac{\sqrt{8+\sqrt{15}}}{2}):=M$. However, using Theorem 1 directly, we obtain $C_{\mathrm{max}}(\rho)=\log_2\left(\frac{3}{2}+\frac{\sqrt{15}}{16}\right):=N$. For $M-N=-0.0160<0$, Theorem 1 cannot be used to  $C_{\mathrm{max}}(\rho)$ directly.

\begin{figure}[h]
\centering
\includegraphics[width=0.45\textwidth]{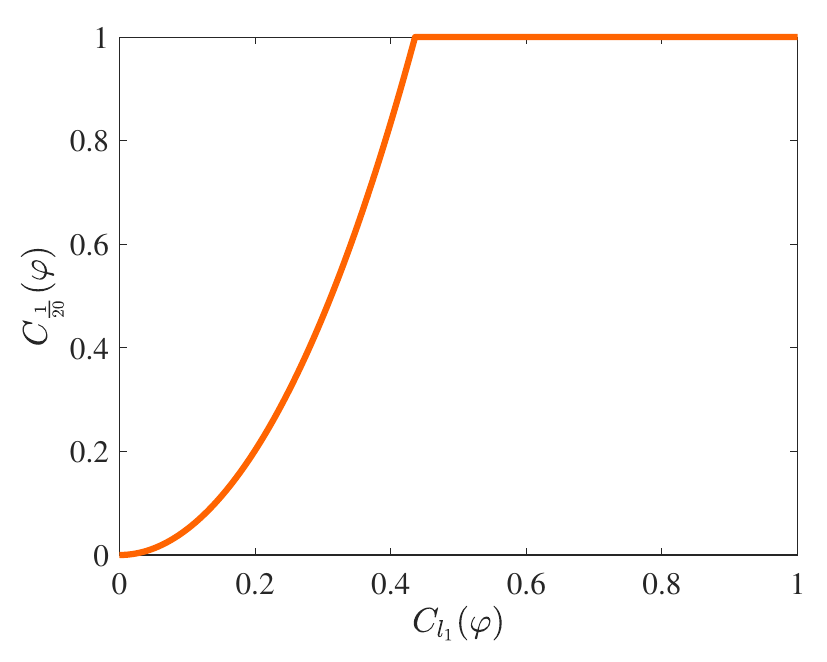}
\caption{The function $C_{\frac{1}{20}}(\varphi)$ of $C_{l_1}(\varphi)$ with $\varphi$ being an one-qubit pure state. $C_\mu(\varphi)$ is neither a convex nor a concave function with respect to $C_{l_1}(\varphi)$. }
\label{fig2}
\end{figure}

Next, let us consider a class of coherence measures $C_\mu(\rho)$ which were presented in Ref. \cite{Du3}. For any given pure state $\ket{\varphi}=c_0\ket{0}+c_1\ket{1}$ with $\abs{c_0}\leq \abs{c_1}$, the coherence measures $C_\mu(\varphi)$ are defined as
\begin{eqnarray}
C_\mu(\varphi):=\bar{f}_\mu(\abs{c_0}^2):=\frac{\abs{c_0}^2}{\mu}\wedge1 \label{pure}
\end{eqnarray}
with $\mu\in[0,1]$. Here, the notation $a\wedge b$ denotes the minimal value between $a$ and $b$ and $C_0(\varphi):=\left\{
  \begin{array}{ll}
    0,  ~~~~~~~~~ \abs{c_0}^2=0\\
    1,  ~~~~~~~~~ \abs{c_0}^2\neq0
  \end{array}\right.$.
With the definition of $C_\mu(\varphi)$ in Eq. (\ref{pure}), they are extended to the mixed states as
\begin{eqnarray}
C_\mu(\rho)=\inf_{\{p_i,\varphi_i\}}\sum_ip_i C_\mu(\varphi_i), \label{measure}
\end{eqnarray}
where the infimum is taken over all the ensembles $\{p_i,\varphi_i\}$ realizing $\rho$. It is straightforward to observe that if $\mu=1$, then $C_\mu(\varphi)$ is a convex function with respect to $C_{l_1}(\varphi)$. If $\mu=0$, then $C_\mu(\varphi)$ is the coherence measure presented in Theorem 2. For values of $\mu$ in the interval $(0,1)$, $C_\mu(\varphi)$ is neither convex nor concave with respect to $C_{l_1}(\varphi)$. Next, we demonstrate that an analytical expression for $C_\mu(\varphi)$ with $0<\mu<1$ cannot be obtained directly using Theorem 1. To see this, let us consider a particular coherence measure $C_\mu(\rho)$ with $\mu=\frac1{20}$. (See Fig. \ref{fig2}).
For the state
\begin{eqnarray}
\rho=
\left(
    \begin{array}{cc}
        \frac{9}{25} & \frac{\sqrt{29}+35}{100}\\
        \frac{\sqrt{29}+35}{100} & \frac{16}{25}\\
    \end{array}
\right),
\end{eqnarray}
there is an ensemble decomposition of $\rho$, $\{p_i,\varphi_i\}$, with
\begin{eqnarray}
&p=\frac{3}{10},~~~\ket{\varphi_1}=\sqrt{\frac{1}{30}}\ket{0}+\sqrt{\frac{29}{30}}\ket{1};\nonumber
\\
&1-p=\frac{7}{10},~~~\ket{\varphi_2}=\frac{1}{\sqrt{2}}\ket{0}+\frac{1}{\sqrt{2}}\ket{1}.
\end{eqnarray}
By direct calculations, we obtain $C_{\frac{1}{20}}(\rho)<1$. However, by using Theorem 1, we obtain $C_{\frac{1}{20}}(\rho)=1$. Thus, Theorem 1 cannot be used to $C_{\frac{1}{20}}(\rho)$ directly.

From the above discussions, we can conclude that if $C_f(\varphi)$ is not a convex function with respect to $C_{l_1}(\varphi)$, Theorem 1 cannot be directly applied. Even though we cannot derive closed expressions for these measures, we can provide an operational interpretation of these measures. Specifically, we can determine the necessary and sufficient conditions for $\rho_1:=p\varphi_1\oplus(1-p)\varphi_2$ to be transformed into $\rho_2:=q\phi_1\oplus(1-q)\phi_2$ by utilizing incoherent operations with $\varphi_i$, $\phi_j$, $i, j=1, 2$ being one-qubit pure states and $0\leq p, q\leq 1$. This leads us to the following theorem:

\emph{Theorem} 3. The state $\rho_1:=p\varphi_1\oplus(1-p)\varphi_2$ can be transformed into $\rho_2:=q\phi_1\oplus(1-q)\phi_2$ using incoherent operations if and only if
\begin{eqnarray}
C_\mu(\rho_1)\geq C_\mu(\rho_2)
\end{eqnarray}
for all $\mu\in[0,1]$, where $C_\mu$ is the coherence measure defined in Eq. (\ref{pure}).

\emph{Proof}.  Let us assume that
\begin{eqnarray}
\ket{\varphi_1}&&=\sqrt{s}\ket{0}+\sqrt{1-s}\ket{1},~\nonumber\\~\ket{\varphi_2}&&=\sqrt{t}\ket{2}+\sqrt{1-t}\ket{3};\nonumber\\
\ket{\phi_1}&&=\sqrt{\theta}\ket{0}+\sqrt{1-\theta}\ket{1},~\nonumber\\~\ket{\phi_2}&&=\sqrt{\tau}\ket{2}+\sqrt{1-\tau}\ket{3}.
\end{eqnarray}
Here, without loss of generality, we assume that $s,t,\theta,\tau\in[0,\frac12]$.
Suppose there is some incoherent operation $\Lambda$ such that $\Lambda(\rho_1)=\rho_2$. We then obtain
\begin{eqnarray}
\Lambda(\varphi_1)&&=q_1\phi_1\oplus(1-q_1)\phi_2\nonumber\\
\Lambda(\varphi_2)&&=q_2\phi_1\oplus(1-q_2)\phi_2\label{transformation}
\end{eqnarray}
with $pq_1+(1-p)q_2=q$. By using a result in Ref. \cite{Du3}, which says that a pure state $\varphi$ can be transformed into a mixed state $\rho$ by using incoherent operations if and only if there is an ensemble $\{p_j,\varphi_j\}$ of $\rho$ satisfying $C_n(\varphi)\geq\sum_jp_jC_n(\varphi_j)$ with $n=1,...,d-1$, we have that
\begin{eqnarray}
s&&\geq q_1\theta+(1-q_1)\tau,\nonumber\\
t&&\geq q_2\theta+(1-q_2)\tau.
\end{eqnarray}

Using the aforementioned claims, we will now demonstrate the if part of the theorem. To achieved this, we only need to consider the following four cases: (i) $s>t\geq\theta\geq\tau$; (ii) $s\geq\theta>t\geq\tau$; (iii) $\theta\geq s>t\geq\tau$; and (iv) $s=t$.

In case (i), it means that $C_\mu(\rho_1)\geq C_\mu(\rho_2)$ for all $\mu\in[0,1]$ and  both $\varphi_1$ and $\varphi_2$ can be transformed into $\phi_1$ and $\phi_2$ with certainty by using incoherent operations. Thus, we only need to choose $q_1=q_2=q$.

In case (ii), we choose $\mu=\theta$. By using the condition $C(p\rho\oplus(1-p)\sigma)=pC(\rho)+(1-p)C(\sigma)$ \cite{Yu}, one obtains that $C_\theta(\rho_1)\geq C_\theta(\rho_2)$ equals to $p\theta+(1-p)t\geq q\theta+(1-q)\tau$. This further implies the condition $q\leq p+(1-p)q_2^m$, where $q_2^m=\frac{t-\tau}{\theta-\tau}$ is the maximal probability of obtaining the state $\phi_1$ from $\varphi_2$ by using incoherent operations. With this condition, we can always achieve the desired transformation in Eq. (\ref{transformation}).

In case (iii), we choose $\mu=1$. Then $C_\theta(\rho_1)\geq C_\theta(\rho_2)$ is $ps+(1-p)t\geq q\theta+(1-q)\tau$. This further implies the conditions $pq_1^m+(1-p)q_2^m\geq q$, where $q_1^m=\frac{s-\tau}{\theta-\tau}$ and $q_2^m=\frac{t-\tau}{\theta-\tau}$ are the maximal probability of obtaining the state $\phi_1$ from $\varphi_1$ and $\varphi_2$ by using incoherent operations, respectively. With these conditions, we can always achieve the desired transformation in Eq. (\ref{transformation}).

In the last case, we choose $\mu=1$. Then the transformation $C_\theta(\rho_1)\geq C_\theta(\rho_2)$  implies $\varphi\to\rho_2$. By using the result that a pure state $\varphi$ can be transformed into a mixed state $\rho$ by using incoherent operations if and only if there is an ensemble $\{p_j,\varphi_j\}$ of $\rho$ satisfying $C_n(\varphi)\geq\sum_jp_jC_n(\varphi_j)$ with $n=1,...,d-1$, there is some incoherent operation achieves the transformation $\varphi\to\rho_2$.

All the above conditions imply that if there is $C_\mu(\rho_1)\geq C_\mu(\rho_2)$ for all $\mu\in[0,1]$, then we can achieve the transformation from $\rho_1$ into $\rho_2$.

Finally, let us consider the form of the incoherent operation in Eq. (\ref{transformation}). Let $\Lambda_1(\varphi_1)=q_1\phi_1\oplus(1-q_1)\phi_2$ and $\Lambda_2(\varphi_2)=q_2\phi_1\oplus(1-q_2)\phi_2$. We may choose the desired transformation
as $\Lambda(\cdot)=\Gamma\circ\Theta(\cdot)$, where $\Theta(\cdot)=\Lambda_1(\cdot)\otimes\ket{0}_a\bra{0}+\Lambda_2(\cdot)\otimes\ket{1}_a\bra{1}$ and $\Gamma(\cdot)=\Tr_a(\cdot)$. It is direct to see that $\Lambda_1(\cdot)$ and $\Lambda_2(\cdot)$ being incoherent operations implies that $\Lambda(\cdot)$ is an incoherent operation. This completes the if part of the theorem.

Let us prove the only if part of the theorem. Suppose there is some incoherent operation $\Lambda$ such that $\Lambda(\rho_1)=\rho_2$. Then, $C(\rho_1)\geq C(\rho_2)$ follows immediately by using the condition (C2), i.e., $C_\mu(\rho_1)\geq C_\mu(\rho_2)$ for all $~\mu\in[0,1]$. ~~~~~~~~~~~~~~~~~~~~~~~~~$\blacksquare$

Before concluding, we would like to make the following remarks regarding Theorem 3.

\emph{Remark} 1. Any coherence measure $C(\rho)$ that satisfies conditions (C1)-(C4) also satisfies the additivity condition \cite{Yu,Liu5}: $C(p\rho\oplus(1-p)\sigma)=pC(\rho)+(1-p)C(\sigma)$. Thus, the coherence properties of the states $p\rho\oplus(1-p)\sigma$ can be reduced to the sum of the coherence properties of $\rho$ and $\sigma$. By using this result, the coherence properties of the state $p\varphi_1\oplus(1-p)\varphi_2$ in Theorem 3 can be reduced to the coherence properties of $\varphi_i$ with $i=1,2$. In other words, let $\rho_i$ be 2-dimensional states, $0\leq p_i\leq 1$, and $\sum_i p_i=1$. Then, $C(\oplus_i p_i\rho_i)=\sum_i p_i C(\rho_i)$. Since the state $\ket{\varphi_2}(\ket{\phi_2})=c_2\ket{2}+c_3\ket{3}$ is an one-qubit state in the space spanned by ${\ket{2},\ket{3}}$, the conditions for the transformation $p\varphi_1\oplus(1-p)\varphi_2\to q\phi_1\oplus(1-q)\phi_2$ via incoherent operations, where $\varphi_i$ and $\phi_j$ $(i, j=1, 2)$ are one-qubit pure states and $0\leq p, q\leq 1$, can be solved by only using the properties of one-qubit states, i.e., $\varphi_i$ and $\phi_j$.

\emph{Remark} 2. The convex roof coherence measures for one-qubit states with $C_f(\varphi)$ being convex with respect to $C_{l_1}(\varphi)$ are not sufficient to characterize the state transformations under incoherent operations as stated in Theorem 3. To see this, let us consider the following two states
\begin{eqnarray}
\rho_1&&=p\varphi_1\oplus(1-p)\varphi_2,\nonumber\\
\rho_2&&=q\phi_1\oplus(1-q)\phi_2,
\end{eqnarray}
where $\ket{\varphi_1}$, $\ket{\varphi_2}$, $\ket{\phi_1}$, and $\ket{\phi_2}$ are
\begin{eqnarray}
&&\ket{\varphi_1}=\frac{1}{\sqrt{2}}\ket{0}+\frac{1}{\sqrt{2}}\ket{1},~\nonumber\\~&&\ket{\varphi_2}=\frac{1}{2}\ket{2}+\frac{\sqrt{3}}{2}\ket{3};\nonumber\\
&&\ket{\phi_1}=\sqrt{\frac{1}{3}}\ket{0}+\sqrt{\frac{2}{3}}\ket{1},~\nonumber\\~&&\ket{\phi_2}=\sqrt{\frac{1}{11}}\ket{2}+\sqrt{\frac{10}{11}}\ket{3},
\end{eqnarray}
and $p=\frac{1}{6},q=\frac{5}{6}$.
Using the result from Theorem 3, we see that it is impossible to transform $\rho_1$ into $\rho_2$ or vice versa using incoherent operations since $C_{\frac{1}{3}}(\rho_1)=\frac{19}{24}<C_{\frac{1}{3}}(\rho_2)=\frac{29}{33}$. However, all coherence measures in Theorem 1 satisfy $C(\rho_1)>C(\rho_2)$ since they are monotonically increasing functions of $C_{l_1}$, and $C_{l_1}(\rho_1)=\frac{2+5\sqrt{3}}{12}>C_{l_1}(\rho_2)=\frac{55\sqrt{2}+3\sqrt{10}}{99}$. Moreover, it is evident that $C_{R}(\rho_1)=C_{R}(\rho_2)=1$. Thus, we can conclude that the coherence measures considered in Theorems 1 and 2 are insufficient to characterize the state transformations in Theorem 3.

\section{Conclusions} \label{V}

In summary, we have studied the closed expression of convex roof coherence measures for one-qubit states. We present the analytical expressions of the convex roof coherence measures, denoted as $C_f(\rho)$, for one-qubit states. In Theorems 1 and 2, we demonstrate that $C_f(\varphi)$ is convex with respect to $C_{l_1}(\varphi)$. Furthermore, we provide operational interpretations of these measures in Corollaries 1 and 2. Although we cannot obtain the analytical expression for non-convex convex roof coherence measures with respect to $C_{l_1}(\varphi)$, we present their operational interpretation by providing the necessary and sufficient conditions for $p\varphi_1\oplus(1-p)\varphi_2\to q\phi_1\oplus(1-q)\phi_2$ by utilizing incoherent operations in Theorem 3. Here, $\varphi_i$, $\phi_j$ with $i, j=1, 2$ are one-qubit pure states and $0\leq p, q\leq 1$. Finally, we anticipate that our results can be applied to other resource theories such as non-classicality \cite{Asboth,Miranowicz,Miranowicz1}.

\begin{acknowledgments}
X.D.C acknowledges support from the National Natural Science Foundation of China through Grant No. 12147155.
C.L.L acknowledges support from the China Postdoctoral Science Foundation Grant No. 2021M690324.
\end{acknowledgments}

\end{document}